\documentclass[aps,prb,twocolumn,showpacs]{revtex4-1}
\usepackage{graphics}
\graphicspath{{figs/}}
\usepackage{subfigure}
\usepackage{epsfig}
\usepackage{epsf,epic}
\usepackage{color}
\usepackage{marvosym }
\usepackage{subfigure}
\usepackage{siunitx}
\usepackage{amsmath}
\usepackage{amssymb}
\usepackage{amsfonts}
\usepackage{wrapfig}
\usepackage{pstricks}
\usepackage{multirow}
\usepackage{pst-node}
\usepackage{extdash}
\usepackage{bm}
\usepackage{dcolumn}
\newcommand{\etal}{\textit{et al.\ }}

\newcommand{\mc}{\multicolumn}
\begin{document}
\title{Quasiparticle self-consistent $GW$ band structures of Mg-IV-N$_2$ compounds:
  the role of semicore $d$ states}
\author{Sai Lyu and Walter R. L. Lambrecht}
\affiliation{Department of Physics, Case Western Reserve University, 10900 Euclid Avenue, Cleveland, Ohio 44106-7079, USA}
\begin{abstract}
  We present  improved band structure calculations of the Mg-IV-N$_2$ compounds
  in the quasiparticle self-consistent $GW$ approximation. Compared to
  previous calculations (Phys. Rev. B {\bf 94}, 125201 (2016)) we here
  include the effects of the Ge-$3d$ and Sn-$4d$ semicore states and find
  that these tend to reduce the band gap significantly. This places the
  band gap of MgSnN$_2$ in the difficult to reach green region of the visible spectrum. The stability of the materials with respect to competing binary compounds is also evaluated
  and details of the valence band maximum manifold splitting and effective masses are provided. 
  
\end{abstract}

\maketitle
The family of II-IV-N$_2$ semiconductors is recently receiving increasing
attention. They are related to the group-III nitrides semiconductors by
replacing the group-III ion by a pair of a group-II and a group-IV ion
in an ordered and octet-rule preserving manner. 
They share many of their opto-electronic properties with the group-III nitrides
which are well known for their applications in light-emitting diodes (LED),
laser diodes, UV-detectors and so on. Another reason for their interest is
that they  many of these compounds provide alternatives to the In and Ga
in the form of sustainable and earth-abundant elements, such as Mg, Si, Sn.
For an overview of these compounds, see Refs.  \onlinecite{Lambrechtbook,Martinez17,ictmc21}. Among this family,  the Mg-IV-N$_2$ compounds were recently studied in a few papers.\cite{Atchara16,Quirk14,Arab2016,Bruls2000,Bruls2001,Bruls1999,Rasander17,Kanchiang2018,Kaewmeechai18}
In particular, quasiparticle self-consistent band structure calculations
of the Mg-IV-N$_2$ were presented by Jaroenjittichai \etal\cite{Atchara16}
The main purpose of the present short note is to point out that
with an improved basis set, including Sn-$4d$ and Ge-$3d$ semicore orbitals
the band gaps of these materials are significantly reduced.
At the same time, we analyze the stability of these materials in
further detail than previously studied by comparison with the
competing binary nitrides.

{\bf Computational approach:} The quasiparticle self-consistent $GW$
method\cite{MvSQSGW,Kotani07}
is used for the band structures in a full-potential linearized muffin-tin
orbital implementation.\cite{Methfessel,Kotani10,questaal} Here, 
$W$ is the screened Coulomb interaction and $G$ the one-electron Green's function and their convolution provides the self-energy $\Sigma$ which incorporates
the dynamical effects of the electron-electron interactions beyond the
density functional theory (DFT).
This approach has been shown to yield accurate band structures, to
better than 0.1 eV accuracy in comparison with experiment, in particular
when the self-energy is scaled by a factor $0.8$. This scaling factor
corrects for  the underestimated screening of the screened Coulomb energy
$W$ in the random phase approximation which neglects electron-hole interaction
diagrams.\cite{Deguchi16,Churna18} Essentially the same approach
was previously used by Jarroenjittichai \etal\cite{Atchara16}.
These results were also found to be in good agreement with hybrid
functional calculations,\cite{Kanchiang2018} although a rather high fraction of exact
exchange (0.5) was required to obtain good agreement with the previous 
QS$GW$ results. Such a large mixing fraction is unusual and already an
indication that there might be a problem. 
The accuracy of the QS$GW$ approach still depends strongly on the
completeness of the basis set and other convergence parameters.
The present calculations improve on the previous work
by using an improved basis set.   As in previous work,
a $3\times3\times3$ {\bf k}-point mesh is used for the calculation of the
$GW$ self-energy and subsequently interpolated to a $6\times6\times6$ mesh
for self-consistent charge density calculations and to the {\bf k}-points
along symmetry lines.
However, while previous work only considered Mg-$2p$ orbitals as
semi-core states, we here include also Sn-$4d$ and Ge-$3d$ as local
orbitals. The levels in fact lie at higher energy (Sn-$4d$ at -1.89 Ry and Ge-$3d$ at -2.15 Ry) than the Mg-$2p$
at (-3.43 Ry) and will be shown to have an important impact in shifting
the valence band maximum (VBM) upwards thereby reducing the gap. 

{\bf Results:}

\begin{figure}
\centering
  \includegraphics[width=8cm]{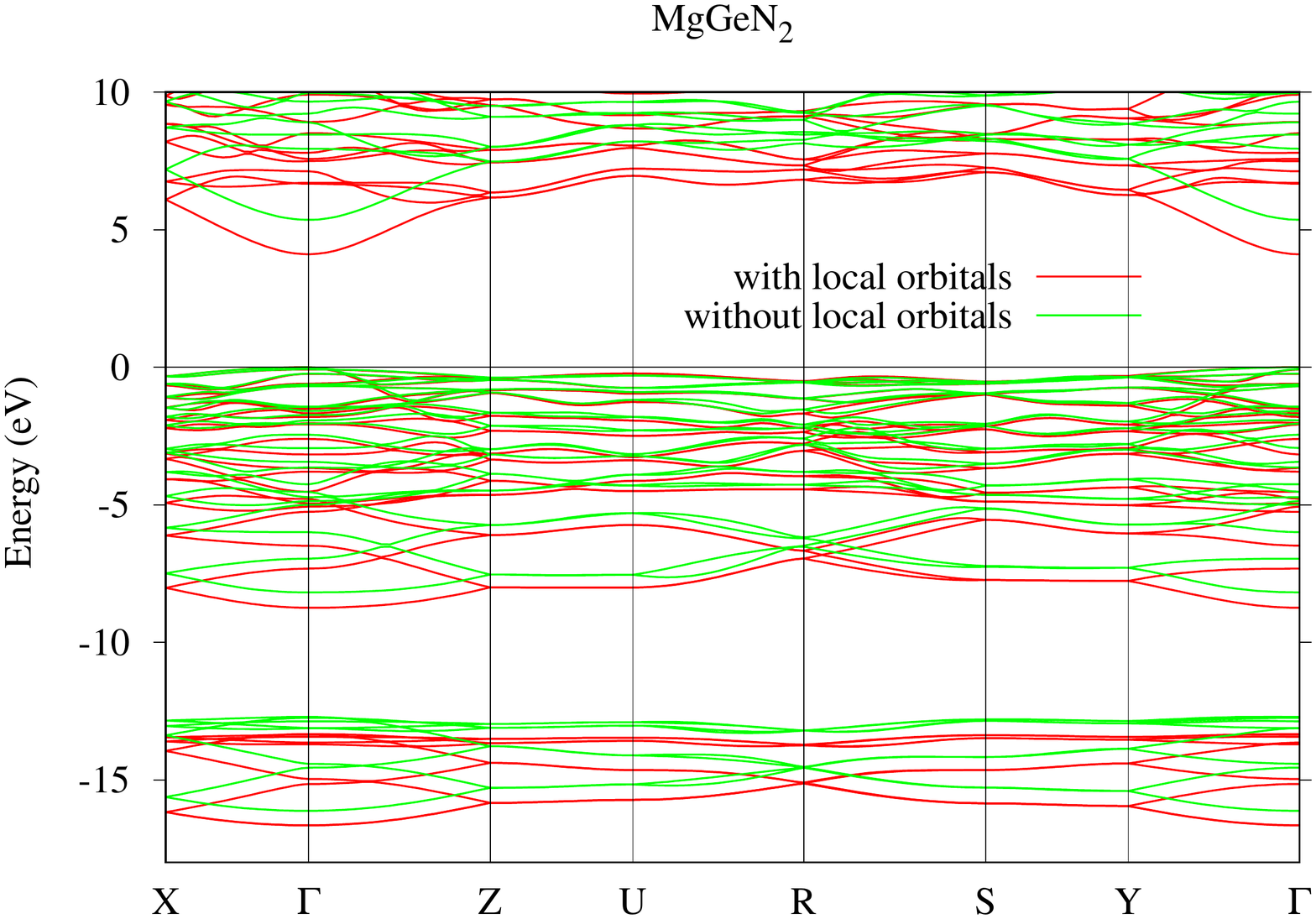} \vskip-1cm
  \includegraphics[width=8cm]{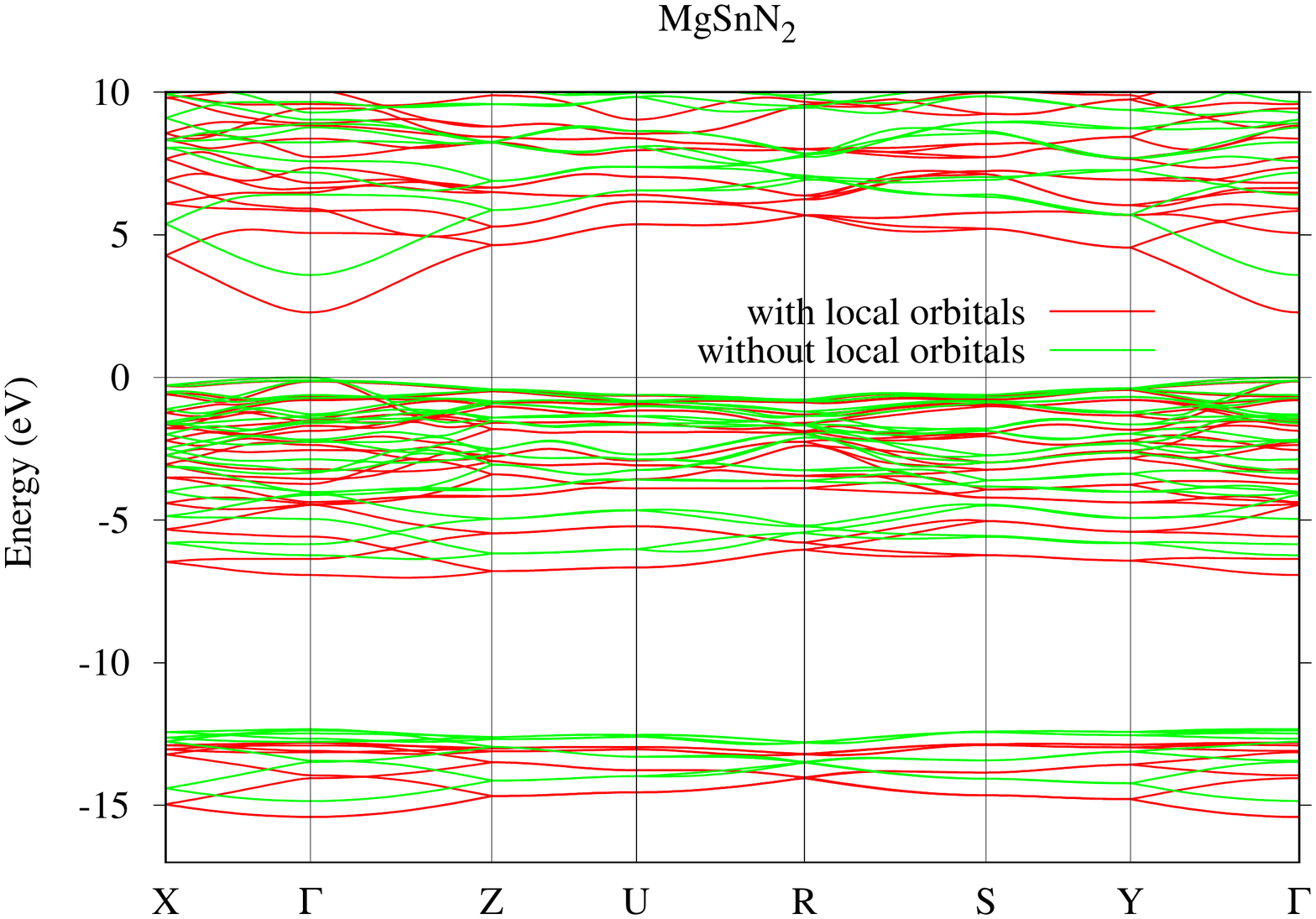}
  \caption{QS$GW$ band structure of MgGeN$_2$ and MgSnN$_2$.\label{figband}}
\end{figure}

In Fig. \ref{figband} we show the band structure of 
MgGeN$_2$ and MgSnN$_2$ both with and without the inclusion of
the local orbitals. The zero reference energy is placed at the valence
band maximum (VBM) in both cases. 
The band effect of these semicore states is clearly visible.
Because the $d$-like core levels interact more strongly with the VBM $p$-like
states than with the CBM $s$-like states, and hybridization of the band
edges with core levels would shift the band edges up, it is
actually the VBM which is shifted up rather than the CBM shifting down. 
The band gaps of MgGeN$_2$ and MgSnN$_2$ are found to be 4.11 eV and 2.28 eV, respectively. Both of them are direct gap semiconductors.

This reduction of the gap is important in particular for MgSnN$_2$.
While previous results suggested a gap close to that of ZnGeN$_2$ and
GaN in the UV region of the spectrum, it is now found to be 2.28 eV
which is in the green region of the spectrum. This is important
because it provides  additional flexibility to design heterostructures
and materials with a band gap in the difficult to reach green
region of the spectrum. The gap deformation potential
$dE_g/d\ln{V}$ can be used to quantify the change in band gap versus the relative change in unit cell volume. From finite-difference calculations, the deformation potential for MgGeN$_2$ is -8.9 eV and for MgSnN$_2$ is -5.4 eV.

\begin{figure}
\centering
  \includegraphics[width=8cm]{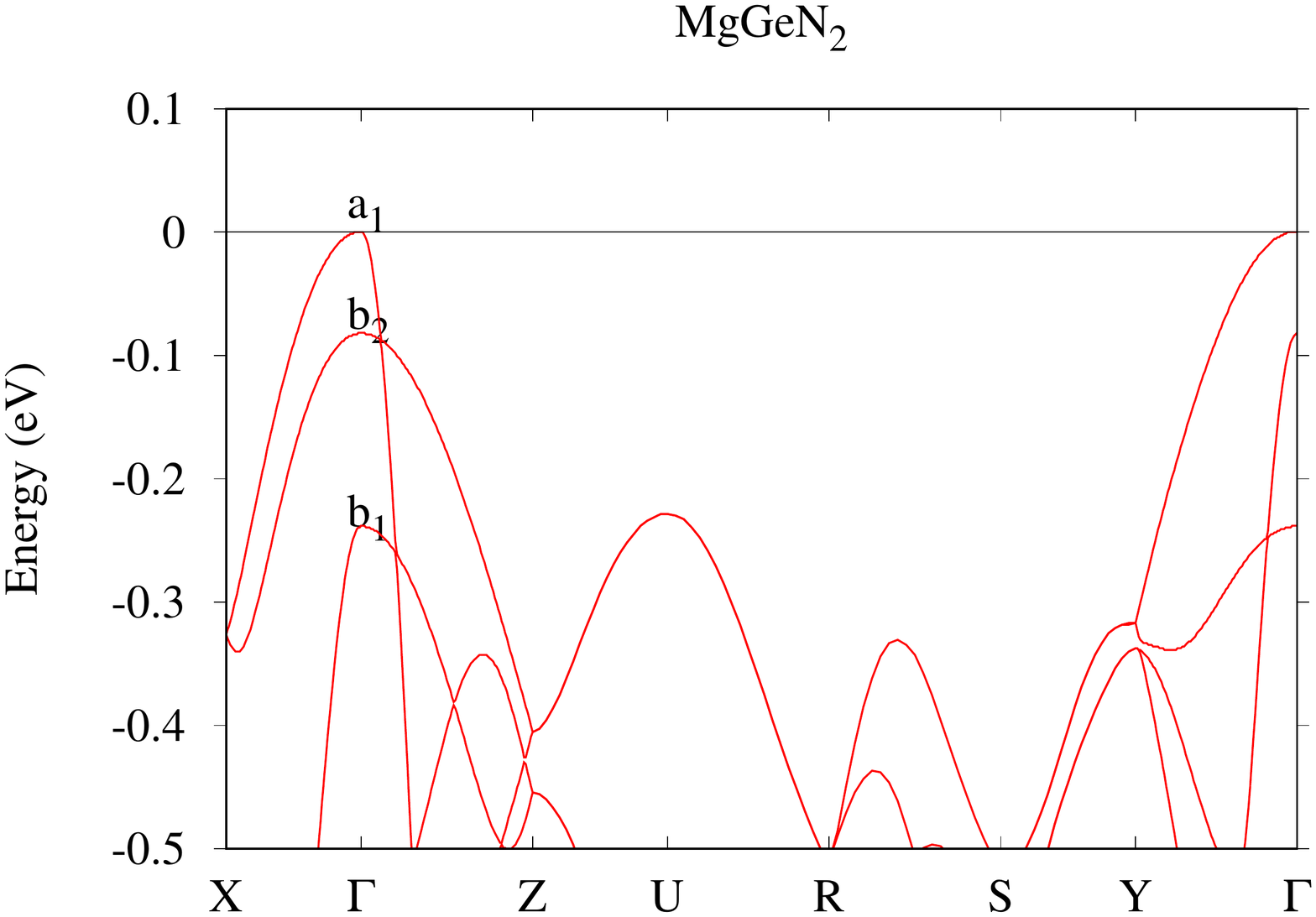} \vskip-1cm
  \includegraphics[width=8cm]{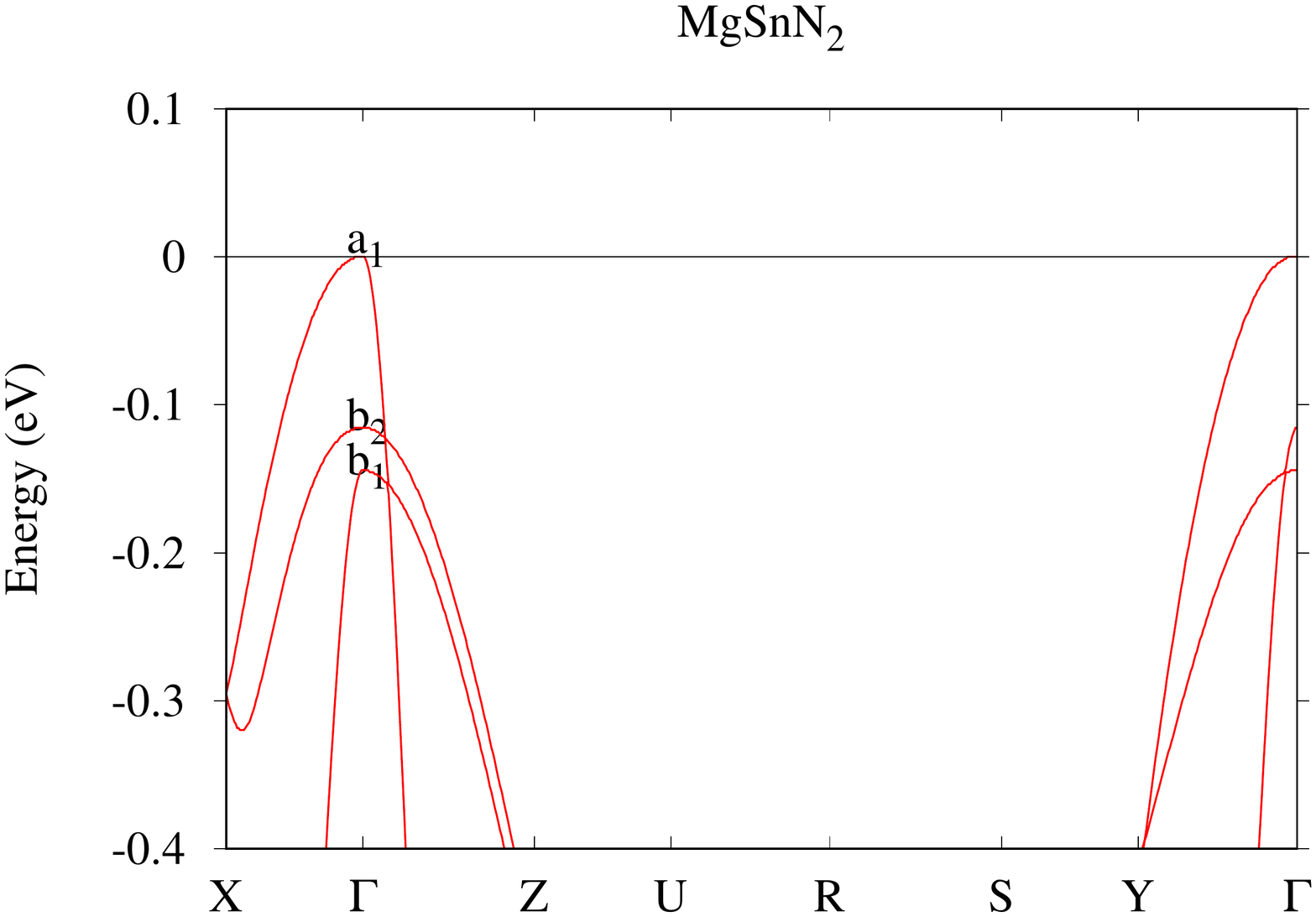}
  \caption{Band structure of MgGeN$_2$ and MgSnN$_2$ near VBM with symmetry labeling at $\Gamma$.\label{figzoom}}
\end{figure}
Next, a zoom in of the band structure near the VBM is shown in
Fig. \ref{figzoom}. The splittings of the VBM, their symmetry labeling
which dictates the allowed optical transitions and the corresponding
effective masses are all summarized in Tables \ref{tabzoomlevel} and \ref{tabmass} providing similar but
updated information as in Ref. \onlinecite{Atchara16}.
\begin{table}
  \caption{Energy levels  in meV at $\Gamma$ relative to the VBM at $\Gamma$
    including their symmetry label.\label{tabzoomlevel}}
  \begin{ruledtabular}
    \begin{tabular}{cdcd}
      \mc{2}{c}{MgGeN$_2$}& \mc{2}{c}{MgSnN$_2$} \\
      Sym.&\mc{1}{c}{$E$}&Sym.&\mc{1}{c}{$E$} \\ \hline
        $a_1$     &0   & $a_1$& 0    \\
        $b_2$     &  -81.6 & $b_2$&-115.7     \\
        $b_1$     &  -238.1   & $b_1$&-144.2     \\
    \end{tabular}
  \end{ruledtabular}
\end{table}

\begin{table}
\centering
\caption{Effective mass  (in the units of the free electron mass $m_{e}$)\label{tabmass}}
 \begin{ruledtabular}
 \begin{tabular}{lcc}
 & MgGeN$_2$ & MgSnN$_2$ \\ \hline
 $m^c_x$ &  0.26       & 0.20     \\
$m^c_y$ &  0.25        &   0.20   \\
$m^c_z$ &   0.24       &   0.19 \\ \\

$m^{a_1}_x$  & 2.46         & 2.60         \\
$m^{a_1}_y$  &  3.53        &  2.81  \\
$m^{a_1}_z$  &  0.22       &  0.18  \\
\\
$m^{b_1}_x$  & 0.29      & 0.21  \\
$m^{b_1}_y$  &  4.56    & 3.54     \\
$m^{b_1}_z$  &   2.54   &2.84       \\
\\

$m^{b_2}_x$  &  2.33       & 2.83  \\
$m^{b_2}_y$  &   0.25       &0.21  \\
$m^{b_2}_z$  &   3.39       &  3.05  \\

  \end{tabular}
  \end{ruledtabular}
\end{table}

\begin{table}
\centering
\caption{Cohesive energies and formation energies (in eV/atom)
  calculated in the GGA-PBE approximation.\label{formation}}
\begin{ruledtabular}
\begin{tabular}{lccc}
    &  This work  & Expt.\footnote{From Gschneider \etal \cite{Gschneider}, except for Mg taken from Kaxiras\cite{kaxiras2003},  N$_2$ taken from Huber and Herzberg\cite{Huber}} &MP\footnote{Material Project\cite{mp}} \\ \hline
Mg  & 1.47 &    1.51               \\
Ge & 3.39 &    3.84 \\
Sn & 2.94 &    3.12 \\
N  & 5.17 &     4.96 \\        \\
Mg$_3$N$_2$  & -0.75 &   &-0.90  \\
Ge$_3$N$_4$   & -0.08  &  &-0.26 \\
Sn$_3$N$_4$   & 0.22 &    &0 \\
MgGeN$_2$   &-0.60   &    &-0.77   \\
MgSnN$_2$ &-0.33    &     &-0.54    \\

\end{tabular}
\end{ruledtabular}
\end{table}

Finally, we discuss the stability of these compounds using GGA-PBE calculations. First we consider the stability of MgGeN$_2$ against binary compounds. With
respect to the reaction,
\begin{equation}
 3\mathrm{MgGeN}_2 \rightarrow \mathrm{Mg}_3\mathrm{N}_2 +\mathrm{Ge}_3\mathrm{N}_4,
\end{equation}
MgGeN$_2$ is found to have a lower energy than
Mg$_3$N$_2$ plus  Ge$_3$N$_4$ by 241 meV/atom. 
Regarding the stability of MgSnN$_2$, the Sn$_3$N$_4$ itself is unstable relative to Sn and N$_2$ which means it lies above the convex hull. So instead of Sn$_3$N$_4$, we need to consider the stability with respect to the following reaction
\begin{equation}
 3\mathrm{MgSnN}_2 \rightarrow \mathrm{Mg}_3\mathrm{N}_2 +3\mathrm{Sn}+ 2\mathrm{N}_2
\end{equation}
We find that MgSnN$_2$ is stable with respect to this reaction and the corresponding reaction  energy is  18 meV/atom. So, in principle both MgGeN$_2$ and MgSnN$_2$ lie on hte convex hull and it should be possible to
synthesize them. 

\acknowledgements{This work was supported by the National Science Foundation,
  Division  of Materials Research
  under grant No. 1533957 and the DMREF program. 
Calculations made use of the High Performance Computing Resource in the Core Facility for Advanced Research Computing at Case Western Reserve University.}
\bibliography{Bib/dft,Bib/abinit,Bib/zgn,Bib/zsn,Bib/msn,Bib/cd4n,Bib/mgn,Bib/pockel,Bib/elastic,Bib/be4n,Bib/gw,Bib/lmto}
\end{document}